\documentclass[aip,graphicx,showpacs,showkeys,sd,reprint,amsmath,amssymb]{revtex4-1}

\usepackage{graphicx}
\usepackage{amsmath}
\usepackage{amssymb}
\usepackage{epstopdf}
\usepackage{dcolumn}
\usepackage{gensymb}
\usepackage{float}
\usepackage{dcolumn}
\usepackage[caption=false,font=footnotesize]{subfig}
\usepackage{color}
\usepackage{natbib}
\begin{document}


\title{Hot carrier redistribution, electron-phonon interaction, and their role in carrier relaxation in thin film halide perovskites}

\author{S. Sourabh}
\affiliation{Department of Physics \& Astronomy, University of Oklahoma, Norman, OK 73019, USA}

\author{V. R. Whiteside}
\affiliation{Department of Physics \& Astronomy, University of Oklahoma, Norman, OK 73019, USA}

\author{Y. Zhai}
\affiliation{National Renewable Energy Laboratory, Golden, CO 80401 USA}

\author{K. Wang}
\affiliation{National Renewable Energy Laboratory, Golden, CO 80401 USA}

\author{V. Yeddu}
\affiliation{School of Materials Science and Engineering, Oklahoma State University, Tulsa, OK 74104, USA}

\author{M. T. Bamidele}
\affiliation{School of Materials Science and Engineering, Oklahoma State University, Tulsa, OK 74104, USA}

\author{D. Y. Kim}
\affiliation{School of Materials Science and Engineering, Oklahoma State University, Tulsa, OK 74104, USA}

\author{M. C. Beard}
\affiliation{National Renewable Energy Laboratory, Golden, CO 80401 USA}

\author{I. R. Sellers}
\thanks{Author to whom correspondence should be addressed:\linebreak
 sellers@ou.edu}
\affiliation{Department of Physics \& Astronomy, University of Oklahoma, Norman, OK 73019, USA}

\date{\today}


\begin{abstract}
Temperature dependent (4 K - 300 K) photoluminescence and transmission spectra are analyzed to study the affect of changing the different components of a perovskite compound, be it A, B, or X. Four different films are compared: FAMAPbSnI$_3$, FAPbI$_3$, FAMAPbI$_3$, and FAPbBr$_3$. The low temperature results highlight the changes that occur especially, underlying ones that are easily masked at room temperature. The overall Stokes shift is of similar magnitude at room temperature for the three Pb only based samples. This is governed by the interaction strength $\Gamma_{LO}$, phonon energy E$_{LO}$, and the exciton binding energy E$_{ex}$. One exception to this behavior is the Sn-based FAMAPbSnI$_3$ film which shows a lack of Stokes shift between the absorption and photoluminescence. However, the strong absorption (more than 100 meV) below the band gap is indicative of an excitonic feature that has a large density of states. Transient absorption measurements confirm the trends observed in continuous wave (CW) measurements, the three Pb only films all show the convolution of an excitonic feature within 20 meV of the band gap as a contributing factor to the photobleach along with a region of high energy PIA. However, the behavior for the Sn-based film is notably different (just as it is in the CW measurements) with an unusual low energy PIA and a lack of high energy PIA. The large unusual low energy PIA is attributed to the large sub band gap absorption observed in the CW transmission/absorption measurements. Notably regardless of interchanging components, the slow cooling of carriers in metal-halide perovsites shows little effect of $\Gamma_{LO}$, E$_{LO}$, and E$_{ex}$. As such, here it is proposed -- while the initial cooling of carriers is attributed to LO phonons, the overall cooling of carriers is dominated by the intrinsic low thermal conductivity of all metal-halide perovskites which limits the dissipation of acoustic phonons in these systems. 
\end{abstract}


\keywords{Excitons, LO phonons, polarons, 2D perovskites, photoluminescence, absorption, transmission} 

\maketitle 

\section{Introduction}


Halide perovskites are ABX$_3$-structured solution-processible semiconductors comprising only earth-abundant elements that can be both fabricated and synthesized at low cost. Here, (compound) `A' refers to a monovalent organic cation such as methylammonium (MA$^{+}$: CH$_3$NH$_3^+$) or formamidinium (FA$^+$), `B' is a divalent metal cation (Pb$^{2+}$, Sn$^{2+}$, or Ge$^{2+}$) and `X' is a halide anion e.g.,  I$^-$, Br$^-$, or Cl$^-$.\cite{Poglitsch1987} By changing the composition of the constituent materials, the electrical and optical properties of the structure can be tailored and controlled.\cite{Mitzi1999} 

Interest in organic-inorganic halide perovskites has resulted in a remarkable increase in solar cell efficiency (2013, 15\% to 25.5\%, 2020) for a material system that first made it on to NREL's best research cell efficiency chart in 2013.\cite{NREL2020} The interest in perovskites has not been limited to solar cell research but extends over a wide range of optoelectronic devices: lasers,\cite{Deschler2014} light emitting diodes,\cite{Tan2014,Stranks2015} and photodetectors.\cite{Deng2015} Additionally, perovskites have shown rather slowed hot carrier cooling\cite{Xing2013,Price2015,Frost2017,Fu2017,Fang2018,Tong2019} in transient absorption measurements spurring considerable interest as possible absorber material for hot carrier solar cells.\cite{Guo2017,Li2019,Lim2020} 

While these slowed carrier cooling rates are measured under short pulsed, high power transient absorption measurements, will those long lifetimes be able to be exploited in a suitably designed hot carrier solar cell? Here, a reexamination of several different bulk perovskites that span a wide energy range, with varying degrees of ionicity and exciton binding energies are considered in order to uncover aspects underlying the unusually slowed hot carrier cooling rates that observed in TA\cite{Manser2014,Price2015,YangNatPhot2015,Anand2016,Yang2017,Frost2017,Fu2017,Fang2018,Tong2019,Hopper2018} or time resolved photoluminescence (TRPL)\cite{Xing2013,Deschler2014} experiments. 

In order to address this question continuous wave (CW) temperature dependent photoluminescence and linear transmission data, as well as, CW power dependent PL are evaluated to determine effective band gaps, exciton binding energies, linewidth broadening, carrier temperature, and Stokes shifts. These trends are subsequently evaluated with respect to phononic properties, ionicity, and therefore possible polaron interactions.

\begin{figure*}[ht]
\centering
\includegraphics [width=1\textwidth] {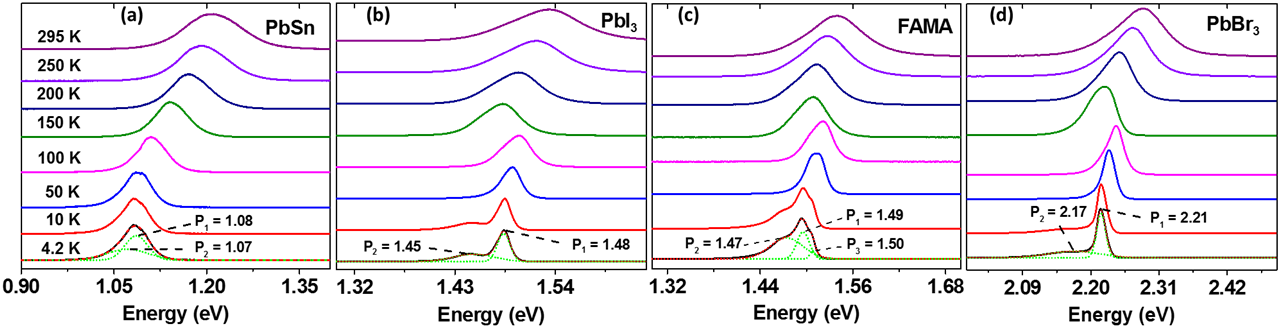}
\caption{Temperature dependent PL from 4.2 K to 295 K (a) FAMAPbSnI$_3$, (b) FAPbI$_3$, (c) FAMAPbI$_3$, and (d) FAPbBr$_3$, respectively.}
\label{normPL}
\end{figure*}
 
\section{Temperature Dependent Photoluminescence}

Figure 1 shows the temperature dependence of the normalized photoluminescence (PL) spectra for four different perovskite thin films (a) FAMAPbSnI$_3$, (b) FAPbI$_3$, (c) FAMAPbI$_3$, and (d) FAPbBr$_3$. All four show a shift to  higher energy as a function of temperature that is typical for metal-halide perovskites due to the p-like conduction band and s-like valence band.\cite{Chang2004,Filip2015} With the addition of methyl ammonium to FAPbI$_3$ there is a slight increase in the PL emission energy (effective band gap) however, the thermally mediated shift in emission remains equivalent, $\sim$50 meV. In general, most parameters extracted in subsequent analysis shows that the addition of methyl ammonium has little affect when directly compared with FAPbI$_3$, indicating the inorganic PbI$_3$ framework dominates the optical properties of these films.         

When considering the PL from the systems under investigation in Figure 1, in all cases, the PL is dominated by a higher energy peak (P1), which is attributed to the exciton emission associated with the band gap, in addition to an additional lower energy peak (P2) and/or an asymmetric low energy tail that is particularly evident at lower temperatures. While this low energy transition (P2) is evident in all the systems studied here, the energy difference with respect to the main PL feature (P1), changes in each case. Specifically, in the case of FAPbBr$_3$ (d) the $\Delta$E$_{p1-p2}$ is $\sim$40 meV (P1 = 2.21 eV; P2 = 2.17 eV). For FAPbI$_3$, $\Delta$E$_{p1-p2}$ is $\sim$10 meV, much smaller than the equivalent Br-based perovskite (Figure 1(d)) while in Sn-based FAMAPbSnI$_3$ (a), $\Delta$E$_{p1-p2}$ is $\sim$10 meV (P1: 1.08 eV; P2: 1.07 eV).

However, in the case of FAMAPbI$_3$ the ground state transition P1 (1.49 eV) consists of an additional peak P3 (1.5 eV), 10 meV higher in energy to P1 (1.49 eV). Although the exact origin of P3 is unknown, cation segregation or the effects of strain are both feasible. The effects of strain and a subsequent reduction in the structural symmetry have been observed previously in FAMAPbI$_3$ which results in a relaxation of optical transition selection rules.\cite{Brauer2020} Such effects would also be expected to lift the ground state degeneracy in FAMAPbI$_3$ resulting in additional fine structure in the system. For FAMAPbI$_3$ (Fig. 1(c)) $\Delta$E$_{p1-p2}$ is $\sim$20-30 meV when considering transition (P3).

In the Pb-based systems, the origin of this lower energy peak (P2) is often attributed to localized or defect mediated emission. If P2 were an intrinsic exciton, it would as the lowest energy transition be expected to dominate the low temperature PL; as well as, to be correlated to the excitonic feature observed in the absorption, neither of which are evident here (see insets to Figure(s) 3(a-c)). 

Here, it is suggested that the origin of this emitting complex (P2) lies in subtle variations in the structural configuration of the different metal-halide perovskite thin films; e.g., the formation of slightly different orthorhombic bond lengths of twist angles that occur at lower temperatures prior to the structural phase transition at $\sim$150 K. Such variations in structure configurations can result from locally different strain fields, which may lift the degeneracy of the band structure in both condensed\cite{Neugebauer1995} and soft matter\cite{Jiao2021} systems or some localized electronegativity differences and the formation of self-trapped excitons\cite{Wong2020} and/or isoelectronic centers.\cite{Brown2017}

As the temperature increases below the structural phase transition at $\sim$150 K, the PL initially shifts to higher energy and a redistribution of carriers between the various transitions is observed. At T $>$ 50 K the PL is dominated by a single feature at the position of transition P1, which is attributed to the intrinsic excitonic contribution at the R-point. While the contribution of P2 steadily decreases with increasing temperature, it continues to contribute to the low energy tail of the emission shifting in tandem with P1 until the phase transition at $\sim$150 K. This redistribution of carriers from P2 to P1 can be understood in terms of the increased thermal energy and the higher density of states of the band edge excitons (P1) -- free carriers -- with respect to P2. Above 50 K, the contribution of this thermal redistribution of carriers along with significant phonon mediated broadening becomes evident. 

In the case of FAMAPbSnI$_3$, the appearance of competing features in the PL appears to be strongly related to its stability and purity, in addition to the structural symmetry of the system. In pristine FAMAPbSnI$_3$ samples the appearance of two features close to the peak of the PL suggests effects similar to that of the pure Pb systems discussed above, which indicates the structure of pristine mixed Pb-Sn alloys may have improved homogeneity and lower intrinsic strain. However, the significantly enhanced low temperature linewidth of the FAMAPbSnI$_3$ samples investigated here ($\sim$70 meV) with respect to the pure Pb samples ($\sim$20-30 meV), likely indicates some inhomogeneity related to either compositional variations\cite{Klug2020} and/or phase segregation. 

Interestingly, similar fine structure was also evident in recent temperature dependent PL for FASnI$_{3}$\cite{Fang2018} indicating that the broad PL could be inherent to Sn-based halide perovskites. The origin of these transitions likely include local and extrinsic variations in structure via degradation induced strain (Sn$^{2+}$ versus Sn$^{4+}$),\cite{Klug2020,Mundt2020,Prasanna2019} local variations in structural symmetry, or electric fields, all of which can affect the degeneracy of the band structure.\cite{Chang2004,Brivio2013,Brauer2020} 

\begin{figure}[htbp]
\includegraphics [width=0.45\textwidth] {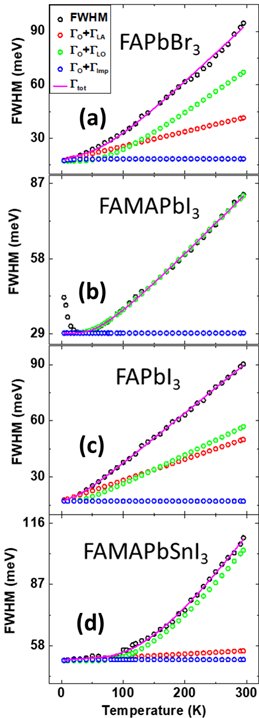}
\caption{FWHM from 4.2 K to 300 K for FAPbBr$_3$ (a), FAMAPbI$_3$ (b), FAPbI$_3$ (c), and FAMAPbSnI$_3$ (d), respectively. The purple solid line is the fit to the data with the contributions of the various coupling parameters as described in Equation (1) shown in red $\Gamma_{0}$ plus $\Gamma_{LA}$, green $\Gamma_{L0}$, and blue $\Gamma_{imp}$.}
\label{FWHM}
\end{figure}

Above, the structural phase transition (T $>$ 150 K), a shift to lower energy is evident in Pb only cases, consistent with the reduced band gap in the tetragonal phase with respect to the low temperature orthorhombic phase.\cite{Stoumpos2013} While the PL is increasingly dominated by the P1 transition, this feature continues to display evidence of a low energy Urbach tail and significant broadening at room temperature providing further evidence for presence of several subtle convoluted and competing complexes at the band edge. But significantly so in the case of FAMAPbSnI$_3$ (see Figure 1(a)), further supporting the presence of alloy fluctuations and segregation in this system.\cite{Klug2020,Mundt2020,Prasanna2019} 

\section{Linewidth Broadening}

To further elucidate the nature of the emission in the metal-halide perovskite investigated here, the temperature dependent PL was evaluated in terms of the thermally activated broadening of the full width half maximum (FWHM). From the FWHM of the temperature dependent PL (see Figure 2) several parameters describing electron-phonon (and impurity related) interactions can be extracted via linewidth broadening analysis. The PL linewidth broadening as function of temperature is given by:\cite{Lee1986,Veliadis1994}   

\footnotesize
\begin{equation}
\Gamma_{tot}(T) = \Gamma_{0} + \Gamma_{LA} T + \frac {\Gamma_{LO}}{[exp\left (\frac {E_{LO}}{k_{B} T} \right)-1]} + \Gamma_{imp} exp\left (-\frac {E_{x}}{k_{B} T} \right),
\end{equation}
\normalsize

\noindent{where, $\Gamma_{0}$, $\Gamma_{LA}$, $\Gamma_{LO}$, and $\Gamma_{imp}$ are the thermal broadening parameters due to the (temperature independent) inhomogeneous broadening (T = 0 K), acoustic (LA) and optical (LO) phonons, and ionized impurities scattering, respectively. The energy of LO-phonons and ionized impurities are denoted $E_{LO}$ and $E_{x}$, respectively.}\cite{Veliadis1994} 

Table 1 lists the most relevant parameters, $\Gamma_{0}$, $\Gamma_{LO}$, and $E_{LO}$. $\Gamma_{LA}$, $\Gamma_{imp}$, and $E_{x}$ are not listed as they are orders of magnitude smaller than the other parameters and do not significantly contribute. While such analysis must be considered with care in the prescence of inhomogeneities and the competition from convoluted transitions, it can be instructive in terms of trends and the qualitative physical behavior.     

\noindent
\begin{tabular}{|c|c|c|c|}
\multicolumn {4} {c} {\textbf{Table 1: Extracted broadening parameters}} \\
\hline
\small Perovskite & \small $\Gamma_{0}$ (meV) & \small $\Gamma_{LO}$ (meV) & \small $E_{LO}$ (meV) \\ \hline
FAMAPbSnI$_3$ & 51 $\pm$ 6 & 204 $\pm$ 20  & 40 $\pm$ 4 \\ \hline
FAPbI$_3$ & 17.2 $\pm$ 2 & 16 $\pm$ 4  & 8.5 $\pm$ 4 \\ \hline
FAMAPbI$_3$ & 29.3 $\pm$ 9 & 43 $\pm$ 4  & 15 $\pm$ 3 \\ \hline
FAPbBr$_3$ & 17.3 $\pm$ 8 & 59 $\pm$ 19  & 20 $\pm$ 10 \\ \hline
\end{tabular}   

The main panels of Figure 2 show the FWHM (solid black circles) with linewidth broadening fitting (red lines with component contributions $\Gamma_{0}$ blue, $\Gamma_{LA}$ green, $\Gamma_{LO}$ yellow, and $\Gamma_{imp}$ purple) as a function of temperature for (a) FAPbBr$_3$, (b) FAMAPbI$_3$, (c) FAPbI$_3$, and (d) FAMAPbSnI$_3$. When assessing the thermal broadening of the PL the dominate peak, P1 is considered only. Moreover, though it is still possible to fit the whole temperature range unlike methylammonium based perovskites\cite{Wright2016} where the fit is limited to temperatures above the crystal reorientation shift, the effects of the structural phase change at $\sim$150 K must also be considered.  

In accordance with the more traditional polar semiconductors such as the III-V systems,\cite{Lee1986,Veliadis1994,Esmaielpour2017} the Fr\"ohlich coupling coefficient ($\Gamma_{LO}$) provides the dominant contribution, which is consistent with the strong polar nature of metal-halide perovskites and the behavior reported in the literature.\cite{Herz2016,Wright2016,Iaru2017,Huang2017} The Fr\"ohlich coupling coefficient ($\Gamma_{LO}$) and the LO phonon energy (E$_{LO}$) of the system determine the carrier thermalization properties and fundamental carrier mobility. In all cases assessed, E$_{LO}$ is consistent with literature values\cite{Herz2016,Wright2016,Iaru2017,Huang2017} and increases as expected with the addition of the lighter element Sn in FAMAPbSnI$_3$. When considering the respective Fr\"ohlich coupling coefficients ($\Gamma_{LO}$), FAPbI$_3$ is less than half the strength of the other metal-halide perovskites. This is consistent with the reduced ionicity of PbI with respect to PbBr, as reflected in the extracted values of FAPbI$_3$ (16 $\pm$ 4 meV) and FAPbBr$_3$ (59 $\pm$ 19 meV), respectively. 

\begin{figure}[htbp]
\centering
\includegraphics [width=0.45\textwidth]{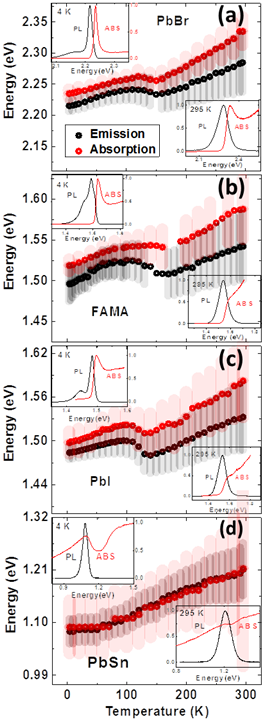}
\caption{(a) Comparison of the temperature dependence of the peak PL energy (black circles) and the energy from the transmission spectra (red circles) for FAPbBr$_3$ (a), FAMAPbI$_3$ (b), FAPbI$_3$ (c), and FAMAPbSnI$_3$ (d) from 4.2 K to 300 K. Shaded regions correspond to the extent of FWHM. The insets show PL (black) and absorbance (red) spectra for 4 K (upper) and 295 K (lower).}
\label{AbsrpEmissEnergy}
\end{figure} 
  
Notably, $\Gamma_{0}$ for the Sn based perovskite is larger than expected (51 $\pm$ 6 meV), more than 3 times larger than FAPbBr$_3$ and FAPbI$_3$. This is attributed to increased defects and low energy tail states associated with Sn-based perovskites and thus not representative of the electron-phonon coupling strength.\cite{Stoumpos2013,Noel2014,Hao2014,Parrott2016} Although recent work has indicated the organic cation may induce additional broadening due to coupling to the inorganic framework in 2D perovskites,\cite{Moral2020} more work is required to make such a conclusion here. However in assessing the increase in FWHM above 150 K the Sn based perovskite has the smallest increase (34 meV), while the Br based perovskite has the largest (55 meV).

Whilst the extracted $\Gamma_{0}$ for FAPbBr$_3$ and FAPbI$_3$ are consistent with literature values,\cite{Herz2016,Wright2016,Iaru2017,Huang2017} the PbBr sample is slightly narrower. This is attributed to the increased separation of P2 from the ground state transition (P1) in FAPbBr$_3$ (see Figure 1(d)), which reduces the convolution of these two transitions and the associated broadening of the dominate emission peak. This is therefore also considered the dominant reason for slightly and progressively larger $\Gamma_{0}$ in the FAPbI$_3$ and FAMAPbI$_3$, respectively, which also display smaller $\Delta$E$_{p1-p2}$ (see discussion regarding Figure 1 above) therefore increasing the convolution of P1 and P2 in these two systems. 

\section{Temperature Dependent Transmission}

To further understand the nature and origin of the transitions observed in the PL for the respective systems, temperature dependent transmission spectroscopy is performed to evaluate the absorption profile of the  thin films, typically a dramatic reduction in transmission is correlated to absorption from a bulk semiconductor's fundamental band gap. It is also possible that below band gap excitonic absorption is exhibited which would correspond to a resonance in the transmission spectrum, with intrinsic complexes providing considerably more absorption than extrinsic transitions created by defects or impurities.\cite{Esmaielpour2020} 
 
The main panels of Figure 3 show both the excitonic absorption feature (red dots, from transmission spectroscopy) and optical transition effective band gap from photoluminescence (black dots) for (a) FAPbBr$_3$, (b) FAMAPbI$_3$, (c) FAPbI$_3$, and (d) FAMAPbSnI$_3$, respectively. The shaded regions correspond to the extent of the FWHM extracted via Gaussian and Voigt fitting of these respective features and reflects the inhomogeneity in the systems with increasing temperature. The upper insets for Figure 3 show a comparison of the absorption and PL spectra at 4 K, while the lower insets show the comparison at 295 K. 

In all of these metal-halide perovskites thin films an excitonic absorption is clearly visible at low temperatures, the onset of which matches the peak of the dominate P1 transition in the PL. Interestingly, the low energy feature P2 is well below that of the main absorption profile, indicating the density of states of the complex or transition P2 is significantly lower than that of the P1, at least in the pure PbI cases. 

Notably, the absorption resonance for FAPbBr$_3$ and FAMAPbSnI$_3$ is easily distinguished even at room temperature. While, for FAPbI$_3$ and FAMAPbI$_3$ it is easily distinguished at 150 K, but that is not the case for room temperature. This is mostly consistent with binding energies that are proportional to the band gap as previously shown in magneto-absorption measurements\cite{Miyata2015,Galkowski2016} and for Wannier-Mott excitons typical of polar semiconductors. Using a fit based on excitonic absorption\cite{Elliott1957} at low temperature where the excitonic feature is more clearly defined, the Br perovskite shows a binding energy comparable to room temperature; whereas, the I based samples are around 160 K or less, these results are similar to those previously reported.\cite{Even2014,Miyata2015} 

What is unusual is the Sn based perovskite showing such a clear absorption resonance even at room temperature with an expected binding energy of only 10 - 15 meV. While there has been fairly wide range of binding energies reported for metal-halide perovskites (see Herz review)\cite{Herz2016} in part due to the indirect methods of determining the binding energies, binding energies in the range 10 - 15 meV are consistent across the literature for FAMAPbSnI$_3$.\cite{Konstantakou2017}   

\begin{figure*}[htpb]
\centering
\includegraphics [width=1\textwidth] {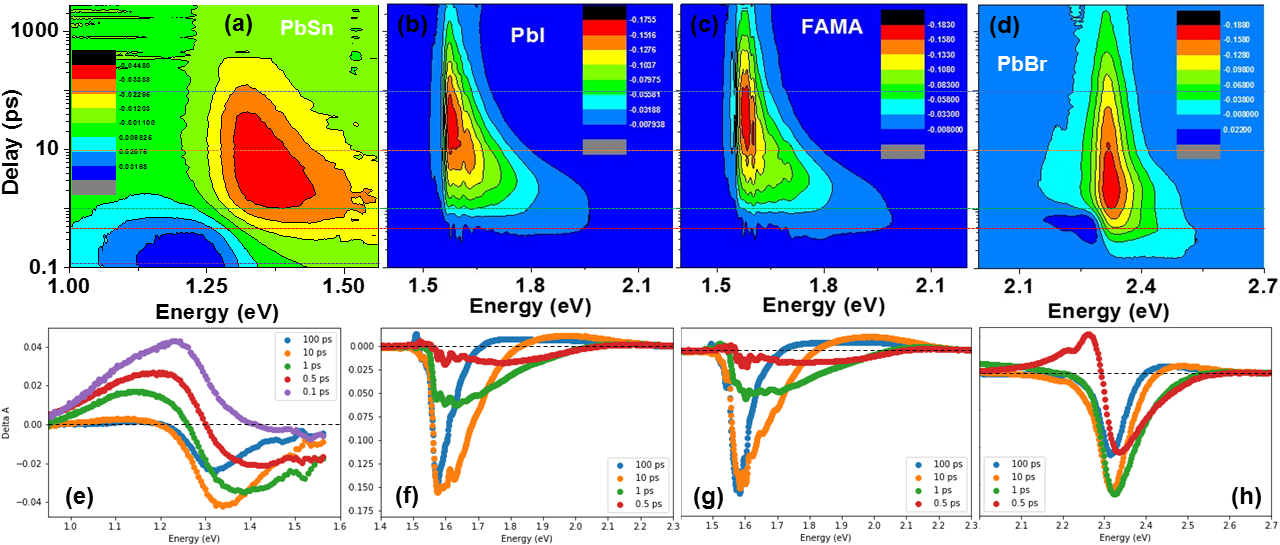}
\caption{Transient absorption measurement of $\Delta$A as a function of energy at room temperature for delay times ranging from 0 to 3000 ps for (a) FAMAPbSnI$_3$ (b) FAPbI$_3$, (c) FAMAPbI$_3$, and (d) FAPbBr$_3$. Transient absorption measurement of $\Delta$A as a function of energy at room temperature for specific (ps) delay times 0.1 purple, 0.5 red , 1 green , 10 orange, and 100 blue for FAMAPbSnI$_3$ (e); 0.5, 1, 10, and 100 for (f) FAPbI$_3$, (g) FAMAPbI$_3$, and (h) FAPbBr$_3$.}
\label{DeltaA}
\end{figure*}

When comparing the emission peak (open black symbols) with the absorptions peak (open black symbols) for FAPbBr$_3$ and FAPbI$_3$ which are the cleanest samples in which only the halide anion is replaced, the respective Stokes shifts at 4.2 K for these two materials are $\sim$20 meV and 12 meV, which reflects both the larger exciton binding energy and ionicity in Br-based systems with respect to I-based materials\cite{Yang2015} and the dynamic nature of carrier-phonon interactions in halide perovskites,\cite{Herz2016,Wright2016,Zhu2015,Frost2017,Miyata2017,Guo2019} and soft matter,\cite{Chang2004,Huang2014} in general.  

The Stokes shift of the mixed cation PbI perovskite, FAMAPbI$_3$, (Figure 3(b)) at 4.2 K is somewhat surprisingly $\sim$20 meV when considering its similarity to FAPbI$_3$, the dominance of the inorganic framework in the electronic and phonic properties, and  the ionicity of FAMAPbI$_3$ with respect to the Br-based FAPbBr$_3$. However, when considering the broadening and strain effects, which result in fine structure around the band edge (P1 and P3 in Figure 1(a)), and the contribution on the PL linewidth (Table 1), a value closer to that of FAPbI$_3$ ($\sim$12 meV) is more probable. 

As the temperature is increased the Stokes shift of FAPbI$_3$ steadily increases to $\sim$50 meV at 295 K. In the case of both FAMAPbI$_3$ and FAPbBr$_3$ the low temperature Stokes shifts remain constant (or slightly reduced) within errors until $\sim$150 K, then they increase to $\sim$50 meV at 295 K. Here, the Stokes shift of the lower energy FAPbI$_3$ appears dominated by interactions with the lattice even at low temperature, while the larger binding energy of FAPbBr$_3$, and FAMAPbI$_3$ (enhanced by local strain) initially inhibit carrier-phonon interactions.

While a Stokes shift in conventional III-V’s systems is indicative of the prescence of sub gap impurity or defect related localization observed by the difference in the associated emission with respect to the continuum -- here, this shift is likely mediated by modulation of the perovskite lattice and polaron interactions that lower the PL energy. The increasing Stokes shift also indicates that the Fr\"ohlich interaction is temperature dependent and increases with temperature. Such behavior has been observed previously and related to the temperature dependence of the dielectric constant and therefore Fr\"ohlich coupling in halide perovskites, which can be attributed to dynamic nature of the lattice and increasing charge contribution at higher temperature in these materials.\cite{Guo2019} 

In the case of the Sn-based FAMAPbSnI$_3$ the difference in peak emission and absorption is observed to be negligible at all temperatures (Figure 3(d)). The origin of this behavior is evident when considering the insets to Figure 3(d), which show the emission and absorption spectra at 4.2 K and 295 K, respectively. The overlap of absorption and photoluminescence has been observed previously 2D Ruddlesden-Popper films\cite{Esmaielpour2020} where there is a large below gap density of states available to absorb. Here, even at low temperature a strong low energy tail is observed in the absorption spectrum, which is related to the sub gap free carrier absorption $\sim$100 meV below the band edge in these nominally p-type materials.\cite{Leitjens2017,Klug2020} It is consistent with the arguments of Wong \textit{et al.}\cite{Wong2021} that shifts in the PL and apparent absorption edge can be observed when there is strong sub band gap absorption that is on the order of \textit{or larger} than k$_B$T. When the relatively large low temperature FWHM ($\sim$70 meV) is also considered (as shown in the shaded area in Figure 3(d)) then it is clear the low binding energy and more importantly ionicity of this system screens any effects of a Stokes shift in this material -- rather than having a low temperature Stokes shift on the order of 10-15 meV, if the behavior is similar to the Pb only films.      

When all of the continuous wave (CW) measurements are taken as a whole, it becomes clear that the three Pb only based films follow similar trends with the exchange of the halide dictating the largest influence on band gap, excitonic binding energy, and linewidth broadening. This is clearly seen in the larger E$_{g}$, E$_{LO}$, and $\Gamma_{LO}$ values for the Br film;  as well as, the more well-defined excitonic absorption feature at room temperature. While there is more structure in the low temperature PL when the FA is replaced with FAMA, the overall behavior between the two Pb-I films is very similar. All three films clearly show that the excitonic feature in the transmission data is within 20 meV of the band edge (indicative of exciton binding energy). Likewise, the films exhibit increased polaron behavior (increasing Stokes shift) with increasing temperature once a threshold has been reached. Here, a larger threshold temperature corresponds to a larger binding energy. 

However, with addition of Sn there is a noticeable difference in behavior which is most obvious in Figure 3 where it is apparent there is neither a Stokes shift nor a shift in energy associated with low temperature crystal phase shift. The below band gap absorption (clearly evident in upper 4 K inset $\sim$100 meV) appears to have the greatest influence on the overall behavior of this PbSn film with regards to CW measurements. With that in mind, an investigation in the dynamics of these films via transient absorption measurements is carried out to compare and contrast these distinct trends across these four films.   

\section{Transient Absorption}

Transient absorption (TA) measurements at high carrier densities have demonstrated slower hot carrier cooling rates than GaAs under identical excitation conditions.\cite{YangNatPhot2015} A typical TA spectra for a metal-halide perovskite shows three distinct regions: photoinduced absorption (PIA) for photon energies less than the band gap at short times due to an exciton stark shift and broadening;\cite{Price2015} a photobleaching due to a decrease of the exciton absorption with phase space filling and bleaching due to the Moss-Burstein shift (state filling); and a high energy region with photoinduced absorption as the carriers cool and the photobleach diminishes at longer delay times, which is associated with carrier density and band gap renormalization.\cite{YangNatPhot2015} 

At first glance one can see considerable differences from the TA spectra in Figure 4 between FAMAPbSnI$_3$ (a), FAPbI$_3$ (b), FAMAPbI$_3$ (c) and FAPbBr$_3$ (d). While all the spectra show some low energy photoinduced absorption at very short delay times and exciton/band gap photobleaching, all but the Sn based film show high energy photoinduced absorption within 250 meV of the photobleach minima. The energy range and strength of the photo-bleach varies considerably at short delay times. The I containing films all have much more extensive bleach (over a broader energy range) at delay times less than 10 ps as compared to the Br film which has a larger band gap and thus a smaller excess pump energy.  

Moreover, there is a blue shift in energy corresponding to the maximum intensity of the exciton photobleach energy which is more discernible in Figure 4(e-h) which show individual time delay traces 0.5, 1, 10, 100 ps (4(e) shows an additional purple trace for 0.1 ps due to the unusual PIA).
At 0.5 ps there is a low energy PIA due to band gap renormalization which is limited in the two PbI
films due to the proximity of the detector cutoff. All the films show the beginning of Moss-Burstein shifted photobleach. The photobleach fully evolves quickly over time ($<$ 10 ps in all but the PbSn film which takes 10's of ps longer to finally reach a steady energy position) red shifting within a few picoseconds to an energy position that is the convolution of the excitonic absorption and the fundamental band gap for the film.

Additionally, by a few picoseconds (2 - 5 ps) there is a high energy PIA region (blue region above the broad photobleach region Figure 4(b,c,d) which starts within 250 meV of the photobleach minima and extends to higher energies which peaks then gradually diminishes) that is carrier density dependent that can be related to the band gap renormalization and the diminishing Moss-Burstein shift as delay time increases -- except in the PbSn film which shows no high energy PIA (Figure 4(a,e)). While the PbSn film does not show as wide an energy window beyond the photobleach minima due to the detector cutoff, it does extend 250 meV away from the photobleach minina for the 10 and 100 picosecond transients (orange and blue traces Figure 4(e)). In all the other films this is a region of high energy PIA however, the PbSn is still within the photobleach region.  

Perhaps what is most striking is that the Sn based sample has the strongest low energy PIA which is consistent with the onset of absorption observed in the transmission measurements (see Figure 3 discussion). Also of note, is that the transmission/absorption exciton energy does not match the TA bleach but rather the low energy PIA. Rather, just the continuum/band gap energy matches well with the TA photobleach energy. In all of the other samples the excitonic absorption feature (see Figure 3) is a major contributor to the TA photobleach. 

Typically, such TA responses are used to extract the carrier temperature relaxation rate that assumes the formation of a Maxwell-Boltzmann distribution at energies larger than the band gap,
 
\footnotesize
\begin{equation}
-\Delta A(E) = A_0(E) [exp\left (\frac {-E}{k_{B} T_c} \right)]
\end{equation}
\normalsize

\noindent {where $\Delta A$ is the change in the transient absorption, $A_0(E)$ is the absorptivity as a function of energy, $k_{B}$ is Boltzmann's constant, $T_c$ represents the carrier temperature which can be extracted from a linear fit of the high energy tail of the transient absorption spectra.} 

Since each of the effects discussed above can exhibit different temporal and carrier density dependencies care must be taken when implementing such analysis as TA spectra comprise a complex convolution of different processes as recently discussed by Yang \textit{et al.}\cite{YangNatPhot2015} and Lim \textit{et al.}\cite{Lim2020} However, these complications are most pronounced near the band edge and thus using the higher energy tail provides the best estimate of the carrier distribution, which enables discussion of the qualitative behavior and temporal dynamics of hot carriers.

Figure 5 shows a comparison of the hot carrier temperature versus time for the FAPbI$_3$, FAMAPbI$_3$, and FAPbBr$_3$ films from the TA spectrum shown in Figure 4(b,c,d). The dynamics are plotted on a logarithmic scale from 1 ps, which is the time at which the ground state bleach and high energy tail are evident in the TA spectra recorded from these films (see Figure 4).

\begin{figure}[htbp]
\centering
\includegraphics [width=.48\textwidth] {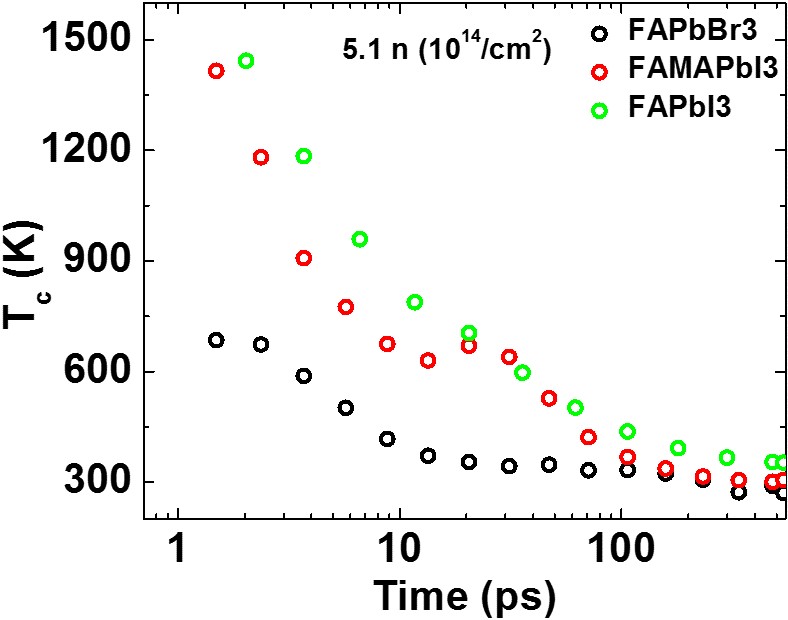}
\caption{T$_c$ as a function of delay time at room temperature for FAPbBr$_3$ (black), FAMAPbI$_3$ (red), and FAPbI$_3$ (green).}
\label{Tc}
\end{figure}
 
\begin{figure*}[htbp]
\centering
\includegraphics [width=1\textwidth] {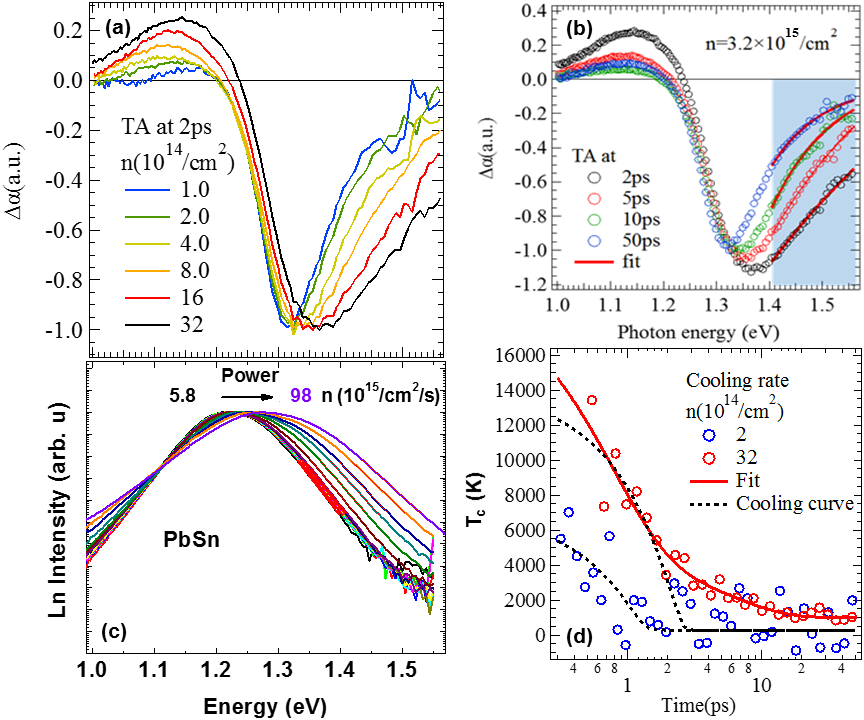}
\caption{a) FAMAPbSnI$_3$ transient absorption spectra of $\Delta$A as a function of energy at room temperature for delay times ranging from 2 to 50 ps b) Carrier density dependence of FAMAPbSnI$_3$ room temperature transient absorption spectra for a delay time of 2 ps c) Room temperature CW carrier density dependence for FAMAPbSnI$_3$ d) Cooling rate curves for carrier densities of 2 and 32x10$^{14}/cm^2$ for FAMAPbSnI$_3$}
\label{DeltaA}
\end{figure*} 

When considering the PbI$_3$ systems with FA (green open circles) and the double cation FAMA (open red circles), the relaxation of hot carriers towards equilibrium (300 K) is similar. This is expected since their band gaps with respect to the excitation energy (400 nm) are similar, resulting in hot carriers of the same energy/temperature at t = 0. Moreover, the ground state bleach and thermalization are dominated by the intrinsic inorganic Pb-halide framework,\cite{YangNatPhot2015} heat capacity, and ionicity of the metal halide, in addition to the available phonon modes\cite{Hopper2018} in the system. The FAPbBr$_3$ displays a lower relative initial carrier temperature since the excess energy created in this system is lower due to its larger band gap with respect to the PbI$_3$ systems.

At t $<$ 1 ps (see Figure 5(a)) the carrier distribution quickly evolves into a Maxwell-Boltzmann distribution via strong carrier-carrier interaction reaching a non-equilibrium (T$_c$ $>$ T$_{L}$) hot carrier distribution followed by (t $>$1 ps) strong Fr\"ohlich coupling and hot LO phonon generation/emission in polar semiconductors. This initial carrier-LO phonon interaction with the lattice results in a fast rate for hot carrier relaxation in the Pb-based perovskites between the $\sim$1 ps to $\sim$10 ps. This is followed by a much slower thermalization rate, typically between a few (2-10) ps to 10-1000 ps which has been observed in several metal-halide perovskites at higher excitation densities ($>$ 10$^{13}cm^{-2}$),\cite{Frost2017,Fu2017,YangNatPhot2015,Lim2020} including the pure-Pb (Figure 5) and Sn-based (Figure 6) perovskites investigated here.

When comparing the relaxation rate of the FAPbBr$_3$ to the PbI$_3$ systems in Figure 5, the relaxation rate of the Br system is greater than that of the I samples. This can be attributed to the increased ionicity of PbBr$_3$, which results in strong electron-phonon interaction and therefore carrier relaxation as observed previously.\cite{Hopper2018} While the initial thermalization of hot carriers in metal-halide perovskites (polar semiconductors, in general) is generally accepted to be the result of carrier - LO phonon interaction, the longer lived regime (t $>$ 10 ps) is less well understood. Recently, this was attributed to carrier heating via Auger processes\cite{Fu2017} due to the high carrier concentrations required to produce this effect.

While this is not discounted here, \textit{since no appreciable effect is seen for different E$_{ex}$ or E$_g$\cite{Yang2015} which would be expected if Auger processes were dominant -- an alternative explanation is provided}. The carrier relaxation pathway of hot carriers in polar semiconductors is dominated initially by Fr\"ohlich coupling and the emission of hot LO-phonons (t $\sim$1 ps). The LO phonons then dissipate via the emission of longitudinal acoustic phonons, t $<$ 10 ps (LA) via either the Klemens (LO = 2LA)\cite{Klemens1966} or Ridley (LO = TO + LA)\cite{Ridley1996} process creating hot LA phonons whose heat is dissipated into the cold acoustic phonon bath and transfer away via thermal conductivity. 

Slowed carrier thermalization has been attributed to decoupling of the dominant Klemens mechanism in metal-halide perovskites,\cite{Fu2017} Ref. \citenum{Fu2017} also shows strong mixing of the TO and LO modes in band structure calculations, suggesting the Ridley channel remains strong. As such, Ridley processes would be expected to dominate hot LO phonon dissipation and efficiently thermalize hot carriers in halide perovskites. An alternative picture is that although hot LO phonons effectively couple to acoustic phonons via the Ridley mechanism,\cite{Ridley1996} that the dissipation of hot LA phonons is perturbed via the low thermal conductivity of halide perovskites.\cite{OnodaYamamuro1990} This is similar to the process in which reabsorption of hot acoustic phonons was proposed previously\cite{Yang2017} to explain the slowed cooling in some metal-halide perovskites. 

Here however, it is suggested that inefficient conduction of heat creates a bottleneck to the dissipation of hot LA phonons into the lattice, reducing the carrier relaxation rate. This is not the same as, a hot LO phonon bottleneck: the inefficient transfer of hot optical phonons into acoustic relaxation pathways. As such, a hot LO phonon bottleneck is inconsistent with the rapid initial relaxation of carriers in TA measurements as shown in Figure 5. 

Such a hot acoustic phonon bottleneck has recently been invoked to describe the slowed carrier thermalization in nanowires\cite{Yong2013,Chen2020nanolett} and InAs/AlAsSb quantum wells\cite{Esmaielpour2016,Esmaielpour2018} where low thermal conductivity localizes acoustic phonons at interfaces; a similar mechanism to that expected in metal-halide perovskite films at grain boundaries and in regions of crystal phase mismatch. This would explain the \textit{ubiquitous long lived carrier relaxation seen across perovskite systems}.
  
Figure 6(a) shows the excitation dependent TA recorded after 2 ps for the FAMAPbSnI$_3$ film. At the lowest excitation density of n $\sim$10$^{14}$ cm$^{-2}$, the ground state bleach is centered at $\sim$1.31 eV. As excitation power is increased, the ground bleach shifts to higher energy with the increasing contribution of high energy tail typically indicative of hot carriers, in addition to an increasing low energy PIA contribution. The PIA is attributed to the increased contribution of sub gap impurity related absorption in these nominally p-type perovskites\cite{Klug2020} (which increases at higher excitation densities ($\sim$10$^{15}$ cm$^{-2}$)); as photoexcited carriers recombine with the background acceptors, the Fermi level shifts into the valence band increasing the effective optical band gap and inducing free carrier absorption. A further consequence of the degeneracy of the Fermi level with the valence band continuum is a broadening of the absorption profile as the valley degeneracy increases.

Figure 6(c) shows the CW power dependent PL which can be correlated to the high power TA  providing further insight into the role of high intensity photoexcitation in the emission and absorption profiles in FAMAPbSnI$_3$. At low power, the peak of the PL at 300 K is centered at $\sim$1.2 eV with a long low energy tail, as is also observed in Figure 1(a). The origin of the PL peak at low power (and low temperature) is related to the transition P2, which in the case of FAMAPbSnI$_3$ has been attributed to the finite contribution of an extrinsic low energy complex related to the unintentional contribution of Sn$^{4+}$ in the lattice resulting in local strain and regions of self-trapped excitons or isoelectronic center like transitions.\cite{Noel2014,Hao2014,HaoJACS2014}

As the power (or temperature – see Figure 1(a)) is increased these states are saturated, the PL shifts to higher energy and is dominated by the intrinsic ground state exciton P1 at $\sim$1.33 eV. With increasing power (n $>$ 10$^{15} cm^{-2}$) the PL shifts to higher energy and broadens, consistent with the shift of the quasi-Fermi level into the valence band and the contribution of the p-type acceptors. 

This behavior is consistent with that of intentionally doped n- and p-type semiconductors such as GaAs and InGaAs, in addition to systems like InN (n-type)\cite{Mahboob2004,Zanato2004} and GaInNAs (p-type)\cite{Kurtz2004,Pavelescu2005,Brown2017} where high levels of unintentional impurities are evident. As such, care must be taken when evaluating such PL for the presence of hot carriers since the bulk of this broadening reflects the background doping distribution in the bands not the presence of non-equilibrium carriers. 

Typically, the carrier temperatures for photoluminescence spectra would be extracted from a linear fit of the photoluminescence following:

\footnotesize
\begin{equation}
I_{PL}(E)= \frac {A(E) E^{2}}{4 \pi^{2} \hbar^{3} c^{2}} [exp\left (\frac {E-\Delta \mu}{k_{B} T_c} \right)-1   ]^{-1}.
\end{equation}
\normalsize

\noindent {Where I is the PL intensity, $A(E)$ is the absorptivity as a function of energy, $k_{B}$ is Boltzmann's constant, $\hbar$ is the reduced Planck's constant, `c' is the speed of light, $T_c$ represents the carrier temperature which can be extracted from the slope of the natural logarithm of the PL spectrum, and $\Delta \mu$ is the chemical potential. This requires that $A(E)$ is constant over the portion of the spectral region to be fit.}

Indeed, such analysis has been the source of intense debate in the III-V hot carrier photovoltaics community,\cite{LeBris2012,Hirst2014,Gibelli2017,Esmaielpour2017} since such analysis implicitly requires a constant pre-factor prior to the exponential term in Equation 3 -- a fixed band gap (similarly for Equation 2). This has been discussed by several works in the III-V community in terms of hot carrier extraction in PL spectra, particularly as related to QWs,\cite{LeBris2012,Hirst2014Yakes,Esmaielpour2017,Nguyen2018} or where impurities or carrier localization are present.\cite{Esmaielpour2018,Whiteside2019} Thus, no such evaluation of the hot carriers can be deduced from Figure 6(c). However, these dynamics are qualitatively evaluated from the high energy tails of the ground state bleach in TA spectrum (See Figure 6(b)), which at short times represents hot carrier relaxation in the $\Gamma_{0}$-valley of the system.

To further elaborate on the subtle differences in the nature and origin of the TA for the FAMAPbSnI$_3$ film the transients are shown for t $<$ 50 ps in Figure 6(b) at an excited carrier density of n = 3.2x10$^{15} cm^{-2}$. These traces are chosen to describe the nature of the absorption and dynamics of the carrier distribution after a non-equilibrium hot carrier distribution has been formed within a few picoseconds. At 2 ps (black open circles) the TA displays a strong photoinduced absorption (PIA) at low energy (1.0 - 1.25 eV) below the band gap of the system which corresponds to the absorption photobleach minima at $\sim$1.3 eV at delay times $\geq$ 50 ps. With increasing time from 2 ps to 50 ps, the PIA reduces with a simultaneous reduction in the position (red shift) and strength of the photobleach. 

The low energy PIA at t $<$ 50 ps is attributed to sub-gap free carrier absorption and the unintentional p-type background doping that exists in Sn-based systems.\cite{Leitjens2017,Klug2020} This is also is also evident through the sub-gap CW absorption shown inset to Figure 3(d) and in the temperature dependent absorption for this system shown in the supplementary information SI:1(a). As such, PIA due to free carriers in the VB can transition to the unoccupied Fermi-level states in the gap, in addition to conventional absorption across the band gap between the occupied and unoccupied valence and conduction bands. Upon photoexcitation at t $>$ 50 ps, photogenerated electrons will occupy the levels available in the gap created by Sn-vacancies. Since this is considerably shorter than the radiative lifetime of system, the absorption edge will decrease to that of the fermi-level position lowering the effective optical band gap. 

Figure 6(d) shows the carrier temperatures versus time extracted from fits to the high energy tail as shown schematically in Figure 6(b) in red. Here, the behavior is presented for low (n = 2x10$^{14} cm^{-2}$) excitation density (open blue circles) and higher (n = 3.2x10$^{15} cm^{-2}$) excitation density (open red circles). The carrier temperature is `hotter' with increasing excitation fluence. This is typically understood in terms of the presence of a hot LO phonon bottleneck, where a large hot phonon density is inefficiently dispersed by the finite acoustic phonons in the system.\cite{Conibeer2008} This effect is further exacerbated by the low thermal conductivity of metal-halide perovskites, which results in an acoustic phonon bottleneck at longer times. 

Further evidence that an LA rather than LO bottleneck is dominant here is the systematic and rapid decrease in carrier temperature at t $<$ 2 ps in Figure 6(d), which is a general property of metal-halide perovskites both, here and elsewhere.\cite{YangNatPhot2015,Fu2017,Lim2020,Monti2020} This indicates that the carrier relaxation in metal-halide perovskites does not follow the simple LO phonon decay process dominant in more traditional bulk semiconductors.\cite{Balkan1989,Rosenwaks1993} But is rather related to the contribution of several complementary processes, the dominant of which arise after LO phonon relaxation, which is likely facilitated in these macroscopically disordered systems by several interface and boundary edge phonons. This will also localize acoustic phonons and reduce thermal conductivity, in addition to the more intrinsic Ridley processes. 

\section{Conclusions}
Temperature dependent (4 K - 300 K) photoluminescence and transmission spectra are analyzed for different components of a metal-halide perovskite thin film, be it A, B, or X. The low temperature results highlight the changes that occur especially, underlying ones that are easily masked at room temperature. For example, the low temperature PL measurements clearly show the additional complexity when FA is replaced with FAMA; both samples that contain FAMA have a more complicated low temperature spectra then either sample that has only FA. While the overall Stokes shift may be of similar magnitude at room temperature for the three Pb only based samples, the pathway to getting there is not. This is governed on one hand, by the interaction strength (the greater the interaction corresponds to larger Stokes shift) $\Gamma_{LO}$ and phonon energy E$_{LO}$; while, on the other hand is the exciton binding energy and available free carriers (a greater exciton binding energy corresponds to fewer free carriers: smaller Stokes shift/delayed temperature response). 

One exception to this behavior is the Sn-based FAMAPbSnI$_3$ film which shows a lack of Stokes shift between the absorption and photoluminescence that may just be obscured due to the rather large low temperature broadening $\Gamma_{0}$ (due to increased inhomogeneities from Sn$^{4+}$). However, the strong absorption (more than 100 meV) below the band gap is indicative of an excitonic feature that has a large density of states. 

Transient absorption measurements confirm the trends observed in CW measurements, the three Pb only films all show the convolution of an excitonic feature within 20 meV of the band gap as a contributing factor to the photobleach along a region of high energy PIA. However, the behavior for the Sn-based film is notably different (just as it is in the CW measurements) with an unusual low energy PIA and a lack of high energy PIA. The large unusual low energy PIA is attributed to the large sub band gap absorption observed in the CW transmission/absorption measurements. Notably, the lack of a clear dependency of hot carrier relaxation on E$_{ex}$, E$_{LO}$, and/or $\Gamma_{LO}$ suggests the slow cooling observed in these systems is the result of a more generic property of metal-halide perovsikes. Here, this is attributed to their low thermal conductivity -- which appears to limit the dissipation of the hot acoustic phonon bath. This results in a phonon bottleneck that serves to block dissipation of hot LO phonons and consequently, the hot carriers in the system.




\section{Acknowledgments}
This research is funded through the Department of Energy EPSCoR Program and the Office of Basic Energy Sciences, Materials Science and Energy Division under Award No.$\#$ DE-SC0019384. M.C.B., Y.Z. and K.W. acknowledge support thru the Center for Hybrid Organic Inorganic Semiconductors for Energy (CHOISE) an Energy Frontier Research Center supported by the Office of Science, Office of Basic Energy Sciences within DOE thru contract number DE-AC36-08GO28308. 


\bibliographystyle{aipnum4-1}
\bibliography{Perov_Stokes_shift_ref}


\end{document}